\documentstyle[12pt,aaspp4]{article}
\def\ga{\mathrel{\hbox{\rlap{\hbox{\lower4pt\hbox{$\sim$}}}\hbox{$>$}}}}
\slugcomment{To appear in ApJ, 476, L63}
\begin{document}
\title{Implications For The Hubble Constant from the First Seven Supernovae at $z \ge 0.35$}
\author{
A.~G.~Kim,\altaffilmark{1,2}
S.~Gabi,\altaffilmark{1,3}
G.~Goldhaber,\altaffilmark{1,2}
D.~E.~Groom,\altaffilmark{1}
I.~M.~Hook,\altaffilmark{2,10}
M.~Y.~Kim,\altaffilmark{1}
J.~C.~Lee,\altaffilmark{1}
C.~R.~Pennypacker,\altaffilmark{1,3}
S.~Perlmutter,\altaffilmark{1,2}
I.~A.~Small,\altaffilmark{1,2}
A.~Goobar,\altaffilmark{4}
R.~Pain,\altaffilmark{5}
R.~S.~Ellis,\altaffilmark{6}
R.~G.~McMahon,\altaffilmark{6}
B.~J.~Boyle,\altaffilmark{7,8}
P.~S.~Bunclark,\altaffilmark{7}
D.~Carter,\altaffilmark{7}
M.~J.~Irwin,\altaffilmark{7}
K.~Glazebrook,\altaffilmark{8}
H.~J.~M.~Newberg,\altaffilmark{9}
A.~V.~Filippenko,\altaffilmark{2,10}
T.~Matheson,\altaffilmark{10}
M.~Dopita,\altaffilmark{11}
and W.~J.~Couch\altaffilmark{12}\\
(The Supernova Cosmology Project)\\
}
\altaffiltext{1}{\vspace{-.1in}E.~O. Lawrence Berkeley National Laboratory,
Berkeley, California 94720; agkim@LBL.gov}
\altaffiltext{2}{\vspace{-.1in}Center for Particle Astrophysics, U.C. Berkeley, California 94720}
\altaffiltext{3}{\vspace{-.1in}Space Sciences Laboratory, U.C. Berkeley, California 94720}
\altaffiltext{4}{\vspace{-.1in}University of Stockholm}
\altaffiltext{5}{\vspace{-.1in}CNRS-IN2P3, University of Paris}
\altaffiltext{6}{\vspace{-.1in}Institute of Astronomy, Cambridge, United Kingdom}
\altaffiltext{7}{\vspace{-.1in}Royal Greenwich Observatory, Cambridge, United Kingdom}
\altaffiltext{8}{\vspace{-.1in}Anglo-Australian Observatory, Sydney, Australia}
\altaffiltext{9}{\vspace{-.1in}Fermilab, Batavia, Illinois 60510}
\altaffiltext{10}{\vspace{-.1in}Department of Astronomy, University of California, Berkeley, California 94720-3411}
\altaffiltext{11}{\vspace{-.1in}Mt. Stromlo and Siding Springs Observatory, Australia}
\altaffiltext{12}{\vspace{-.1in}University of New South Wales, Sydney, Australia}
\begin{abstract}
The Supernova Cosmology Project has discovered over twenty-eight
supernovae (SNe)
at $0.35 <z< 0.65$ 
in an ongoing program that uses
Type Ia SNe
as high-redshift distance indicators.
Here we present measurements of the
ratio between the locally observed and global Hubble constants, $H_0^L/H_0^G$,
based on the first 7 SNe of this high-redshift
data set compared with 18
SNe at $z \le 0.1$ from the Cal\'{a}n/Tololo survey.
If $\Omega_M \leq 1$, then light-curve-width corrected SN magnitudes
yield
$H_0^L/H_0^G < 1.10$ ($95\%$ confidence level)
in both a $\Lambda=0$ and a flat universe.
The analysis using the SNe Ia
as standard candles without
a light-curve-width correction yields similar results.
These results rule out 
the hypothesis that
the discrepant ages of the Universe derived
from globular clusters and recent
measurements of the Hubble constant are attributable to
a locally underdense bubble.
Using the Cepheid-distance-calibrated absolute magnitudes for SNe Ia
of \cite{sandage:1996},
we can also measure the global Hubble constant, $H_0^G$.
If $\Omega_M \ge 0.2$, we find that 
$H_0^G < 70{\rm~km~s^{-1}~Mpc^{-1}}$ in a $\Lambda=0$ universe and
$H_0^G < 78{\rm~km~s^{-1}~Mpc^{-1}}$ in a flat universe,
correcting the distant and local SN apparent magnitudes
for light curve width.
Lower results for $H_0^G$ are obtained if the magnitudes are not
width corrected.
\end{abstract}


\section{Introduction}

Some of the recent Cepheid measurements in galaxy clusters
suggest a high value of the Hubble constant,
$69 \le H_0 \le 87 \rm{~km~s^{-1} Mpc^{-1}}$
(e.g., Pierce et al. 1994;
\nocite{pi:hubble}
Freedman et al. 1994;
\nocite{fr:hubble}
Tanvir et al. 1995).
\nocite{tanvir:hubble}
However, if the cosmological constant is zero,
such a large Hubble constant
predicts an age of the Universe that is lower than the calculated
ages of globular clusters (Bolte \& Hogan 1995).
\nocite{bo:1995} To account for
this discrepancy, it has been proposed that the locally
(redshift $z \le 0.05$) observed Hubble
constant, $H_0^L$, is actually higher than the global ($z > 0.3$) Hubble constant,
$H_0^G$ (Bartlett et al. 1995).\nocite{ba:h30}
Alternatively, it may be that these Hubble constant measurements 
lie on the tail of their statistical and systematic error distributions.
We use our first sample of seven $z>0.35$ Type Ia supernovae (SNe Ia) to
address both these possibilities, first by directly comparing our SN Ia
sample
with one lying within the local Hubble flow to determine
the ratio of $H_0^L$ to $H_0^G$, and then by
using our sample (the first SNe observed in this redshift regime) together 
with SN Ia absolute magnitude calibrations to
determine the value of $H_0^G$.

The possibility that $H_0^L/H_0^G \ne 1$ has arisen in the context of
the observation of peculiar velocity fields (e.g.,
de Vaucouleurs 1958; Dressler et al. 1987; Lynden-Bell et al. 1988).
%.
%The results of \cite{de:1958},
%\cite{dr:stream}, and \cite{ly:stream}
%suggest that local measurements of the Hubble constant may differ
%from the mean global value.
Simulations of
\cite{tu:hubble} have shown that measured Hubble constants depend on
the observer location and the depth of observations.
Previous work by \cite{lauer:1992} 
has
constrained deviations from uniform Hubble
flow to be $\Delta H_0/H_0 < 0.07$
at $0.01\le z \le 0.05$ using brightest cluster galaxies
as a distance indicator.
The same sample of galaxies 
shows evidence for a 
peculiar motion of $689 {\rm~km~s^{-1}}$
with respect to the cosmic background radiation
(Lauer \& Postman 1994),\nocite{lauer:1994} although 
\cite{ri:1995lg} argue that SNe Ia at
similar redshifts do not support this conclusion.
We thus must still examine the possibility of
a large scale ($z \ge 0.05$) peculiar
velocity flow affecting all the local 
$H_0$ measurements.

The Supernova Cosmology Project has discovered over twenty-eight
SNe in the
redshift range $0.35 < z < 0.65$ in a systematic search
(Perlmutter et al. 1994; 1995).\nocite{sne1994} \nocite{sne1995}
%This new sample of SNe is a potentially valuable tool for cosmology:
The peak magnitudes of these high-redshift candles, when
compared with the peak magnitudes of local SNe,
can yield measurements
of the cosmological parameters $\Omega_M$ and $\Lambda$ 
(Goobar \& Perlmutter 1995; Perlmutter et al. 1996b).
\nocite{goo:lambda}\nocite{pe:1996}
This
calculation implicitly assumes that the local SN
calibrators lie within the global cosmological flow; i.e., that we do not
live in a local bubble where peculiar velocities
appreciably bias the observed value of the Hubble
constant.
In this paper we take an alternative approach,
leaving $\Omega_M$ and $\Lambda$
as free parameters and using our high redshift SNe Ia
to measure
%(in \S2)
the ratio between the locally observed Hubble constant
and the global Hubble constant, $H_0^L/H_0^G$.

We also use our SNe Ia to obtain
a measurement of the Hubble constant.
%(\S3).
This can be compared
to the
other SN-based measurements
which range from ${\rm~57~km~s^{-1} Mpc^{-1}}$ (Sandage et al. 1996)
\nocite{sandage:1996}
to $\sim 66 {\rm~km~s^{-1} Mpc^{-1}}$
(Hamuy et al. 1995; Riess, Press, \& Kirshner 1996), and to the
\nocite{ha:hubble}\nocite{ri:lcs2}
above mentioned Cepheid methods
that connect distances in a sequence
from a single galaxy, to the core of its cluster, and then to the Coma cluster.

%\section{The Distant SN Ia Sample}
%
%We have developed and implemented a systematic search for SNe
%at redshifts $\ga 0.3$.  The first seven SNe from this search
%were discovered between
%1992 and 1994, on the rising side
%of their light curves.
We use the first seven SNe from our search.
A detailed description of our search methodology,
the telescopes used,
the photometric and spectroscopic
data compiled for each event, light curve analysis,
and a study of possible systematic uncertainties, are given in
Perlmutter et al. (1996a,b).\nocite{pe:1996} \nocite{pe:aig}
%Table~\ref{SN} summarizes the properties of our SNe that
%are relevant to this paper:
Specifically,  we use the redshift
as measured from the host galaxy spectrum,
the best fit K-corrected $B$ peak magnitude after our galaxy
extinction correction
$m_B=m_R-K_{BR}-A_R$, the value of $\Delta m_{15}$
(Phillips 1993)\nocite{ph:1993},
and $m_B$ after correction to 
the Leibundgut template $m_B^{\{1.1\}}$
using the relation of \cite{hamuy:1996} as discussed in \S 2.
%For a detailed discussion of the determination of these numbers
%and of the evidence for the
%Type Ia classification, see \cite{pe:1996}.
%For this paper, we take these seven SNe to be
%Type Ia with no evolutionary effects.

\section{The determination of $H_0^L/H_0^G$} 

In order to use SNe Ia as a cosmological candle, we first
must calibrate their luminosities.
If the absolute distance to a SN is known, such as from Cepheids
in the same galaxy, we can obtain the absolute magnitude
$M$ from the apparent magnitude $m$.  More commonly, we can only
measure the redshift and an apparent magnitude.  From these quantities
we can obtain
the intercept ${\cal M}$ of the magnitude axis
of the Hubble relationship,
$m=5\log{cz}+{\cal M}$.
(Following the notation of Perlmutter et al. 1996b,
the script variable indicates a quantity that can be measured
without knowing $H_0$ or the absolute distance.)
These two independent observables 
are related at low redshifts by the relation
%\begin{equation}
${\cal M}=M-5\log{H_0}+25$,
%\label{mandm}
%\end{equation}
where $H_0$ is in units of ${\rm~km~s^{-1} Mpc^{-1}}$.
We call ${\cal M}$ the ``Hubble intercept'' magnitude
%or the ``magnitude zero point'' 
and we use it instead of $M$
when studying relative values of the Hubble constant.

Progress has been made in determining both $M$ and ${\cal M}$
using nearby supernovae.
The Cal\'{a}n/Tololo Supernova Search has discovered and measured a large
sample of SNe Ia within the local Hubble flow, from which 
a Hubble diagram with narrow magnitude dispersion
can be produced and the Hubble intercept
${\cal M}$ fitted.
The sample includes 18 SNe discovered
no later than 5 days past maximum with
redshifts ranging from $3.6<\log{(cz)}< 4.5$.  (Of these, half are 
objects with $cz > 15000 {\rm~km~s^{-1}}$, beyond the distance of the
\cite{lauer:1994}
galaxy cluster sample.
However, they have magnitudes consistent with
the SNe Ia at lower redshift.)
Using these 18 supernovae, \cite{hamuy:1996} find
%\begin{equation}
${\cal M}_B=-3.17 \pm 0.03$
%\label{hint}
%\end{equation}
with rms dispersion $\sigma=0.26$ mag.

Recent advances have led to a more detailed understanding of
SNe Ia.
%:
%there is now compelling evidence that SNe Ia are not all identical
%but still represent a family of objects.
A correlation between peak magnitude and light-curve shape has been
found: \cite{ph:1993}
and \cite{ha:hubble}
parameterize the light curve
with the $B$-band magnitude difference between peak and 15 days after peak
($\Delta m_{15}$) while Riess, Press, \& Kirshner (1995a; 1996)\nocite{ri:lcs}
\nocite{ri:lcs2}
characterize the light-curve shape by the amount ($\Delta$)
of a correction
template needed to be added to a
\cite{le:sup} template to
get a best $\chi ^2$ fit.
These parameterizations within the Type Ia class,
as well as those involving spectral features
(Fisher et al. 1995; Nugent et al. 1995),
\nocite{fi:hubble}\nocite{nu:1995} may make
it possible to use the SNe Ia as
a ``calibrated'' candle with $B$ magnitude dispersions of $< 0.2$ mag.

The \cite{hamuy:1996} sample gives a linear relation between
$\Delta m_{15}$ and the magnitude of the supernova,
which can be expressed in terms of the Hubble intercept:
\begin{equation}
{\cal M}_{B,corr}=(0.86 \pm 0.21)(\Delta m_{15}-1.1) - (3.32 \pm 0.05).
\label{hinthamuy}
\end{equation}
This relation is used to ``correct'' observed
SN magnitudes to a $\Delta m_{15}=1.1$ standard template magnitude,
${\cal M}_B^{\{1.1\}}$.
Applying this correction
reduces the rms dispersion to $\sigma=0.17$ mag for the observed
range of $\Delta m_{15}$, between 0.8 and 1.75 mag.

Not all SN Ia samples show a strong correlation
between light-curve shape and peak magnitude.
\cite{sandage:1996} cite the apparent lack of such a relation
in the Cepheid calibrated SNe Ia to argue for the use of uncorrected
``Branch-normal'' SNe Ia -- that is
SNe with high quality data that pass a simple $\bv$ color selection
or
have no spectroscopic peculiarities.
This subset of SNe Ia has a low dispersion in $B$ magnitude of $\sim 0.3$
mag (Vaughan et al. 1995).
%For this paper
We therefore calculate $H_0^L/H_0^G$
using both light-curve-shape corrected and uncorrected magnitudes.

To measure $H_0^L/H_0^G$, we relate the locally derived values of the
Hubble intercept and the high-redshift observed
magnitudes
using the standard Friedmann-Lema\^{\i}tre
cosmology.  The expected
peak magnitude of a SNe Ia at redshift $z$ 
is a function of the mass density of the universe $\Omega_M$ and the
normalized cosmological constant $\Omega_\Lambda \equiv \Lambda/(3H_0^2)$ :
\begin{eqnarray}
m_R(z) & = & M_B+5\log({\cal D}_L(z;\Omega_M,\Omega_\Lambda))+K_{BR}+25-5\log{H_0^G} \label{answer2} \\ 
 &= & {\cal M}_B+5\log({\cal D}_L(z;\Omega_M,\Omega_\Lambda))+K_{BR}+5\log(H_0^L/H_0^G),
\label{answer}
\end{eqnarray}
(e.g., Peebles 1993; Goobar \& Perlmutter 1995)\nocite{peebles:1993}
where
$K_{BR}$ is the K correction relating $B$ magnitudes of nearby
SNe with
$R$ magnitudes of distant objects (Kim, Goobar, \& Perlmutter 1996) and
${\cal M}_B$ is measured in the local Hubble flow.
\nocite{kim:kcorr}
Here we use ${\cal D}_L$, the ``Hubble-constant-free'' part of the
luminosity distance, $d_L$:
\begin{equation}
{\cal D}_L(z;\Omega_M,\Omega_\Lambda) \equiv d_LH_0=
\frac{c(1 + z)}{\sqrt{|\kappa| }} \; \; \; {\cal S}\! \left (
\sqrt{|{\kappa}| } \int_0^{z} \left [(1+z^\prime)^2(1+\Omega_M z^\prime)-
z^\prime (2+z^\prime ) \Omega_\Lambda \right]^{-\frac{1}{2}} dz^\prime
\right ),
\label{R}
\end{equation}
where for $\Omega_M + \Omega_\Lambda > 1$,
${\cal S}(x)$ is defined as $\sin(x)$ and $\kappa
= 1 - \Omega_M - \Omega_\Lambda $;
for $\Omega_M + \Omega_\Lambda < 1$, ${\cal S}(x) =
{\rm sinh}(x)$ and $\kappa$ as above; and for $\Omega_M + \Omega_\Lambda = 1$,
${\cal S}(x) = x$ and
$\kappa =1$, where $c$ is the speed of light in units of
${\rm km~s^{-1}}$.
We use ${\cal M}_{\rm B}=-3.17 \pm 0.03$
%from Equation~\ref{hint}
for uncorrected magnitudes, and 
${\cal M}_{\rm B,corr}$ from Equation~\ref{hinthamuy}
for light-curve-shape corrected magnitudes. 
For the high-redshift corrected SN magnitudes, $m_R$,
we use only the five SNe whose light-curve widths lie
within the range
($0.8< \Delta m_{15} < 1.75$ mag)
of the local SNe from which the correlation
was obtained: SN1994G, SN1994H, SN1994al, SN1994am, and SN1994an.
The full sample of 7 high-redshift SNe
is used when no correction is applied.

\begin{figure}[tbph]
  \plotone{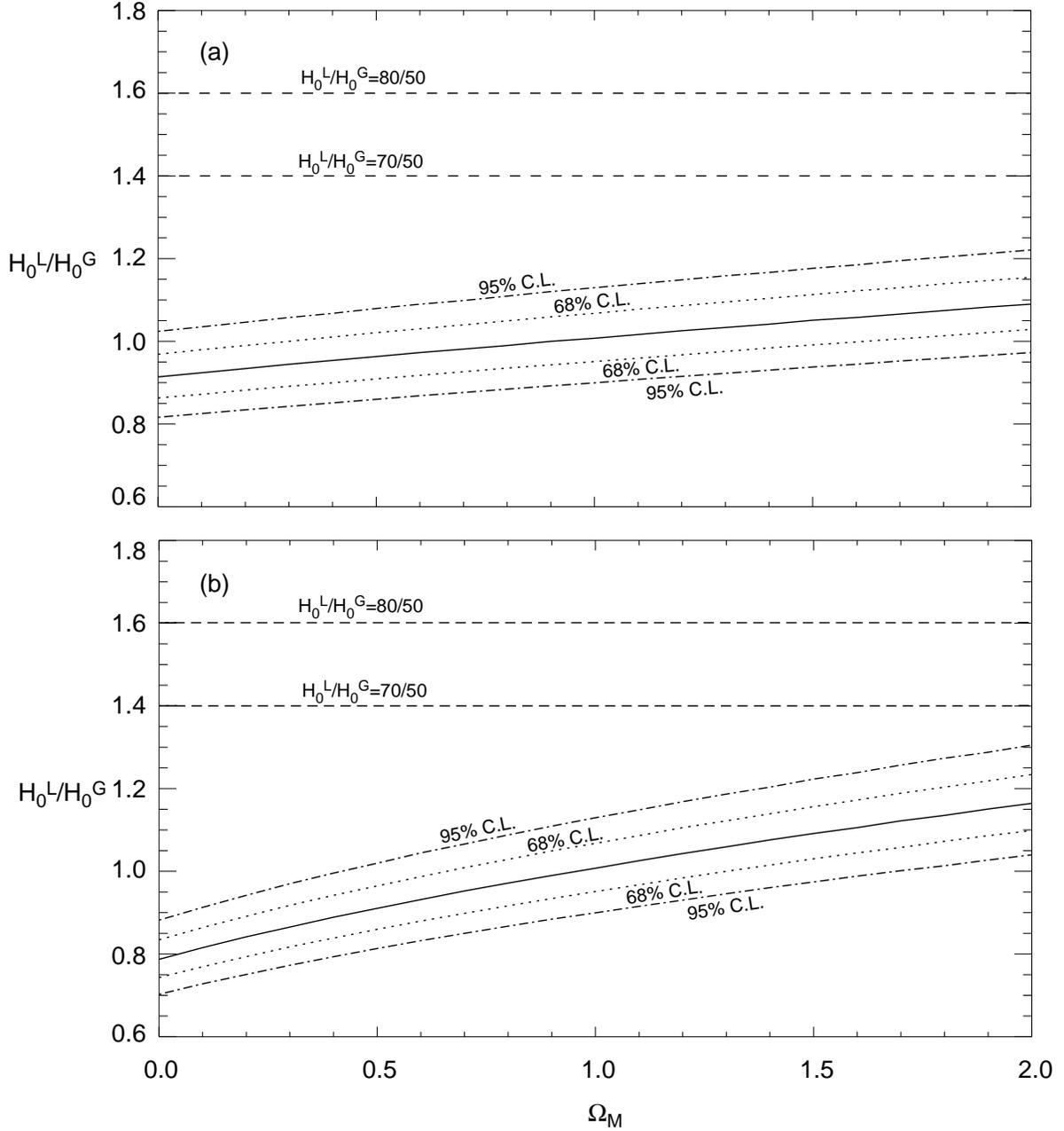}
  \caption[figurecaption]{The best fit $H_0^L/H_0^G$ with $68\%$ (short
  dashes) and $95\%$ (dot-dashes)
  error range for each value of $\Omega_M$ in an (a) $\Lambda=0$ universe
  and (b) flat universe,
  using the five light-curve corrected SN magnitudes.
  (These are the results from a single parameter fit; the uncertainties
  are calculated for each value of $\Omega_M$.)}
  \label{fig1}
\end{figure}

Figure~\ref{fig1}(a) shows the best fit values of $H_0^L/H_0^G$ and
the associated confidence interval curves for a range of $\Omega_M$
in a $\Lambda=0$ universe, based on the light-curve-width corrected
SN magnitudes.
Figure~\ref{fig1}(b) is the same plot as Figure~\ref{fig1}(a) but for
the case of a flat universe ($\Omega_M+\Omega_\Lambda=1$).
Note that the best fit curve is more steeply
sloped than for the $\Lambda=0$ case, increasing the variation
in $H_0^L/H_0^G$ in this $\Omega_M$ range.
(The same plots for the seven
uncorrected magnitudes are almost identical on this scale.)
Also plotted for reference are the ratios of representative high and low
Hubble constant values.

Table~\ref{bounds} has the single-tailed $95\%$ confidence limits (C.L.) for
$H_0^L/H_0^G$ in
$\Lambda=0$ and flat universes using corrected and uncorrected
SN magnitudes.  The lower bounds are calculated at $\Omega_M=0$ where
$H_0^L/H_0^G$ is a minimum.
The value of
$H_0^L/H_0^G$ increases monotonically with respect to $\Omega_M$, so
to obtain an upper limit we choose an upper bound of $\Omega_M \le 2$.
Note that the tabulated numbers
are one-tailed $95\%$ C.L. limits,
unlike the two-tailed confidence intervals given in Figure~\ref{fig1}.

As a cross check, we calculate results for
``Branch-normal'' SNe Ia with
uncorrected magnitudes.  Only SN1994G and SN1994an
are confirmed ``Branch-normal'' based on their color or spectrum and the
results obtained from them are
statistically consistent with
those of the full sample.  In a $\Lambda=0$ universe, we obtain
the limits $H_0^L/H_0^G > 0.79$ and $H_0^L/H_0^G < 1.27$,
while for a flat universe we obtain
$H_0^L/H_0^G > 0.68$ and $H_0^L/H_0^G < 1.35$.

\begin{figure}[tbph]
  \plotone{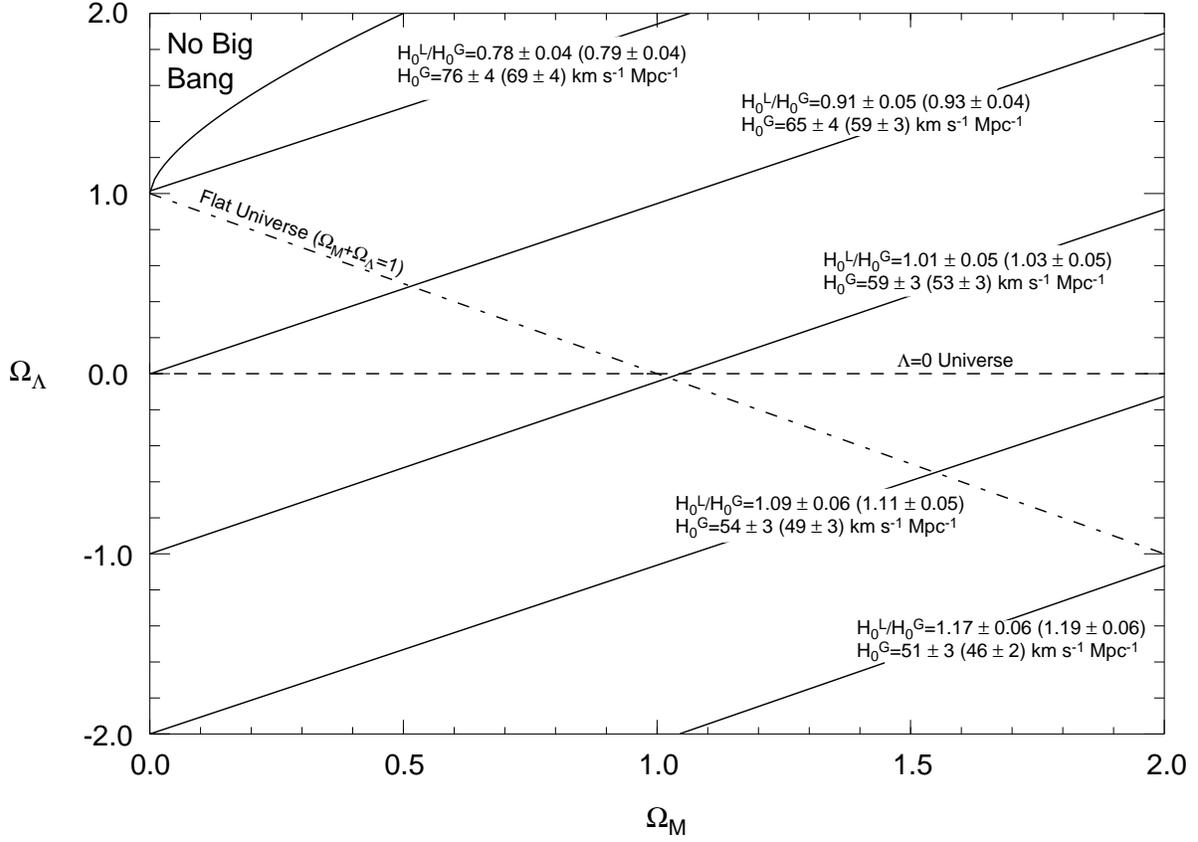}
  \caption[figurecaption]{
    The solid lines show
    contours of constant $H_0^L/H_0^G$ and $H_0^G$ when
    $\Omega_M$ and $\Omega_\Lambda$ are fixed.  They are labeled with their
    value and associated uncertainties
    based on the five corrected SN
    magnitudes.  The values of $H_0^L/H_0^G$ and $H_0^G$ derived from
    the seven uncorrected supernova magnitudes are given in parentheses
    for the approximately corresponding contour.}
  \label{fig3}
\end{figure}

Generally we can calculate $H_0^L/H_0^G$ for any $\Omega_M - \Omega_\Lambda$
pair using Equation~\ref{answer};
we have performed this calculation for a grid of
points in the plane from
$0 \le \Omega_M \le 2$ and $-2 \le \Omega_\Lambda \le 2$
using the five corrected SN magnitudes.
Figure~\ref{fig3} shows curves
of constant $H_0^L/H_0^G$
and associated uncertainties
on the $\Omega_M - \Omega_\Lambda$ plane
as determined from these calculations.
Given in parentheses
on the same plot are the $H_0^L/H_0^G$ values for the same contours
based on calculations from all seven uncorrected SN magnitudes.
(The corrected and uncorrected 
contours do not have the exact same shape in
the $\Omega_M-\Omega_\Lambda$ plane,
%due to their redshift dependence.
%However,
but their deviations are small within the scale of our plot
and in comparison with
our error bars.)
%We thus present the results from both scenarios in a 
%single plot.)
Within the $\Omega_M-\Omega_\Lambda$ region plotted, 
$H_0^L/H_0^G=70/50=1.4$ is excluded
to $\gg 99\%$ confidence.
This limit can still be lower if 
independent lower limits of the age of the Universe and
$\Omega_\Lambda$ are included.

%\clearpage

In \cite{pe:1996}, we discuss the potential errors due to Malmquist
bias and host galaxy extinction.  The bounds on those errors are small
enough not to affect our results.

\section{The Hubble Constant}
The measurement of the global Hubble constant $H_0^G$,
as opposed to the ratio of Hubble constants $H_0^L/H_0^G$,
requires knowledge of the absolute magnitude $M$.
The high resolution of the Hubble Space Telescope has made possible
the discovery of Cepheids and measurement of their 
light curves in galaxies that have hosted
well-observed SNe Ia.
\nocite{sandage:1996}\nocite{saha:1994}\nocite{saha:1995}
To date, six galaxy distances
have been calculated to determine the peak absolute magnitudes of seven
SNe,
giving a weighted mean of
%.  The weighted mean of these SN magnitudes is given in
%\cite{sandage:1996} as
%\begin{equation}
$M_B=-19.47 \pm 0.07 {\rm~mag}$
%\label{snmag}
%\end{equation}
with a dispersion $\sigma=0.16$ mag
(Sandage et al. 1996;
Saha et al. 1994, 1995).

Six of the
seven SNe have a $\Delta m_{15}$ measurement
%(Sandage et al. 1996),\nocite{sandage:1996}
from which we calculate the 
weighted mean of the peak absolute magnitude of SNe Ia corrected
to $\Delta m_{15}=1.1$ mag.
Using the $\Delta m_{15}$ vs. magnitude
relation of Equation~\ref{hinthamuy},
we find
%\begin{equation}
$M_B^{\{1.1\}}=-19.45 \pm 0.07 {\rm~mag}$
%\label{snmaghamuy}
%\end{equation}
with $\sigma=0.14$ mag.

There is some debate on whether these SNe have been properly
extinction-corrected and weighted.
For example, \cite{ri:lcs2} use the correction template
method to conclude that
SN1972E is significantly
extinguished by its host galaxy.  It has also been noted that
the SNe measured with photographic
plates give magnitudes that are systematically brighter than ones measured
photoelectrically.  Therefore, although we use all seven
(six for the $\Delta m_{15}$ corrected) SNe for our main results,
we
include for comparison results from
the \cite{ri:lcs2}
analysis of the three SNe with photoelectric data that yields
%\begin{equation}
$M_{B,\Delta=0}=-19.36 \pm 0.1 {\rm~mag}$
%\label{snmagriess}
%\end{equation}
for a $\Delta=0$ Leibundgut template supernova.

Inserting the absolute magnitude $M_B=-19.47 \pm 0.07 {\rm~mag}$
and the $\Delta m_{15}$--corrected
absolute magnitude $M_B^{\{1.1\}}=-19.45 \pm 0.07 {\rm~mag}$
%of Equations~\ref{snmag}
%and \ref{snmaghamuy}
into Equation~\ref{answer2},
we obtain useful upper bounds of the ``global''
Hubble constant $H_0^G$, which are
listed in Table~\ref{bounds}.
The bounds are calculated at $\Omega_M=0$
for $\Lambda=0$ universes and flat universes because $H_0^G$ decreases
with increasing $\Omega_M$.  If we take $\Omega_M \ge 0.2$ we obtain
even tighter limits, also given in Table~\ref{bounds}.
Figure~\ref{fig3} shows $H_0^G$
in the most general case, for different values of
$\Omega_M$ and $\Omega_\Lambda$.
Note that a value of $H_0$ as high
as $80 {\rm~km~s^{-1} Mpc^{-1}}$ is only found for large values of
$\Omega_\Lambda$ and low $\Omega_M$.
As a cross check, we again calculate our results for uncorrected
``Branch-normal'' supernovae.
We then find $H_0^G < 70{\rm~km~s^{-1} Mpc^{-1}}$
in a $\Lambda=0$ universe and $H_0^G < 82{\rm~km~s^{-1} Mpc^{-1}}$
in a flat universe.

\section{Conclusions}

The measurement of cosmological distances using high-redshift
SNe with locally-calibrated standard
candles sets a limit on the
differences between the local and
global Hubble constants.
From our analysis, it is clear that these data are inconsistent
with scenarios that use a local bubble with high $H_0^L$ that differs
greatly from
$H_0^G$.  We also obtain an upper limit for the Hubble constant
that is consistent with many of the other current measurements.
However,
%tighter
limits that disagree with higher $H_0^G$ measurements may
be obtained with independent upper limits on $\Omega_\Lambda$.

The SN Ia absolute
magnitude calibrations are still subject to debate and may have
systematic errors larger than the statistical ones given above,
so it is important to ask how robust our results are.
An uncertainty in the absolute calibration $\delta m$
in magnitudes
propagates into $\delta H_0/H_0 \approx \delta m$.  A 0.09 mag
difference in the magnitude calibrations, such as the one between the 
$\Delta m_{15}$-corrected absolute magnitudes for six SNe,
$M_B^{\{1.1\}}=-19.45 \pm 0.07 {\rm~mag}$,
%(Equation~\ref{snmaghamuy})
and that of \cite{ri:lcs2},
$M_{B,\Delta=0}=-19.36 \pm 0.1 {\rm~mag}$,
%with their
%extinction corrections for three of the SNe
%(Equation~\ref{snmagriess}),
will
produce a $10 \%$ change in either $H_0^G$ or $H_0^L/H_0^G$.

There is little difference between magnitude
corrected and uncorrected results for the ratio $H_0^L/H_0^G$,
but there is a systematic difference for $H_0^G$ itself, as seen
in Table~\ref{bounds}.
This is because
both the light-curve-width
distribution and the width-magnitude relation of our high-redshift sample
are similar to the distribution and relation
of the \cite{hamuy:1996} sample but not to those
of the
\cite{sandage:1996} sample.
Although these differences may be due to selection effects,
the small number statistics of the Cepheid-calibrated SN sample
can also produce fluctuations that account
for the differences.

In \cite{pe:1996} we calculated $\Omega_M$ and $\Omega_\Lambda$
setting $H_0^L$ equal to $H_0^G$, whereas in this paper
we have discussed the measurement of 
$H_0^L/H_0^G$ while leaving $\Omega_M$ and $\Omega_\Lambda$ as free parameters.
Ideally one would like to measure both sets of quantities simultaneously.
(This problem has been discussed in Wu, Qin, \& Fang 1996.)\nocite{wu:1996}
Filling in a Hubble diagram
with measurements of spatially well-distributed SNe
should make it possible to decouple local and global streaming motions
by showing redshift dependent deviations from the standard model,
and allow one to measure $\Omega_M$ and $\Omega_\Lambda$
independently of local peculiar flows.
Using SNe from
redshift regimes with no evidence of flows, we can simultaneously fit
$H_0^G$, $\Omega_M$, and $\Omega_\Lambda$ using Equation~\ref{answer2},
thus producing a measurement of the Hubble constant.
Our current data set,
which spans from $0.35 < z < 0.5$, shows no sign of peculiar flows
but needs higher statistics and more complete spatial coverage to confirm
this result.

This work was supported in part by the National Science Foundation
(ADT-88909616, AST-9417213) and the
U.~S. Department of Energy (DE-AC03-76SF000098). 
\clearpage
\begin{deluxetable}{cccccc}
\tablecaption{The 95\% One-Tailed Confidence Levels for $H^L_0/H^G_0$
\label{bounds}}
\tablehead{ \colhead{} & \colhead{} & \multicolumn{2}{c}{$H_0^L/H_0^G$} &
\multicolumn{2}{c}{$H_0^G$ Upper limit}\\ \cline{3-4}
\colhead{} & \colhead{} & \colhead{Lower Limit} & \colhead{Upper Limit} &
\multicolumn{2}{c}{${\rm(km~s^{-1}~Mpc^{-1})}$}\\
\cline{5-6} \colhead{} & \colhead{} & \colhead{($\Omega_M \ge 0$)} & \colhead{($\Omega_M \le 2$)} &
\colhead{$\Omega_M \ge 0$} & \colhead{$\Omega_M \ge 0.2$}}
\startdata
$\Lambda=0$ & Corrected & $ > 0.83$ & $ < 1.20$ & $ < 71$ & $ < 70$\\
 & Uncorrected & $ > 0.86$ & $ < 1.21$ & $ < 65$ & $ < 63$\\\hline
$\Omega_M+\Omega_\Lambda=1$ & Corrected & $ > 0.77$ & $ < 1.27$ & $ < 83$ & $ < 78$\\
 & Uncorrected & $ > 0.75$ & $ < 1.30$ & $ < 78$ & $ < 70$\\
\enddata
\end{deluxetable}
\clearpage
\end{document}